\documentclass[english,aps,pra,reprint,noshowpacs]{revtex4-1}   
\usepackage[T1]{fontenc}	
\usepackage[latin9]{inputenc}	
\usepackage{geometry}		
\usepackage{graphicx}
\usepackage[above,below]{placeins}	
\usepackage{times}
\begin{document}

\title{Assessment of critical thinking in physics labs}
\author{Cole Walsh}
\affiliation{Laboratory of Atomic and Solid State Physics, Cornell University, Ithaca, NY, 14853}
\author{Katherine N. Quinn}
\author{N. G. Holmes}
\affiliation{Laboratory of Atomic and Solid State Physics, Cornell University, Ithaca, NY, 14853}
\begin{abstract}
Despite the significant amount of time undergraduate students spend in introductory physics labs, there is little consensus on instructional goals and accepted diagnostic assessments for these labs. In response to these issues, we have developed the Physics Lab Inventory of Critical thinking (PLIC) to assess students' proficiency with critical thinking in a physics lab context. Specifically, the PLIC aims to evaluate students' skills in making sense of data, variability, models, and experimental methods and to assess the effectiveness of lab courses at developing these skills. Here, we discuss the statistical and expert validation of this instrument using 2681 valid student responses collected from 12 institutions during the 2017-2018 academic year. As a part of our validation, we address the impact of lab design on student performance on the instrument.
\end{abstract}

\maketitle

\section{Introduction}

\vspace{-0.2cm}
The goals of instruction in undergraduate physics labs typically involve reinforcing student conceptual knowledge of topics introduced in lecture and, to a lesser extent, teaching students how to work with experimental equipment~\citep{millar2004role}. Though research has shown that the instructional goals of labs are highly debated and vary across disciplines and institutions~\citep{hofstein2004laboratory}, it has been shown that traditional labs do not measurably affect student learning of physics content~\citep{holmes2017value}. There are, however, many important skills that a lab setting uniquely allows students to learn~\citep{holmes2015teaching}. Particularly, proficiency in making sense of data, variability, models, and experimental methods are all skills that are best developed in physics labs~\citep{etkina2006using}. These skills make up a recently endorsed set of recommended learning goals for undergraduate physics lab courses by the American Association of Physics Teachers~\citep{kozminski2014aapt}.

With ongoing laboratory course transformations now occurring at multiple institutions in an effort to meet these new instructional goals, there is an increasing need for validated ways to measure student acquisition of these skills and few validated assessments exist. In response to this, we have developed the Physics Lab Inventory of Critical thinking (PLIC). Here, we demonstrate the concurrent validity~\citep{wilcox2016open}---that is, how consistent performance is with certain expected results---of the PLIC. We expect that either from instruction or selection effects, performance on the PLIC should be higher with greater physics maturity of the respondent. We define physics maturity by the level of the lab course that students were enrolled in when they took the PLIC. We also address the impact of lab courses that have undergone specific transformations to meet the goals outlined above, which we refer to as Structured Quantitative Inquiry labs or SQIlabs~\citep{holmes2015teaching}, on PLIC performance. This work is part of the ongoing validation and reliability assessment of the PLIC following steps laid out in Refs.~\citep{adams2011development} and~\citep{madsen2017resource}. Other validation studies of the PLIC have been published here~\citep{HolmesPERC2015} and here~\citep{quinn2018interview}.

\vspace{-0.4cm}
\section{The PLIC}

\vspace{-0.2cm}
The PLIC is a 4-page online survey that uses a combination of Likert-style and `select all that apply' questions. The Likert-style questions ask respondents to evaluate how well data agrees with a model or how well a particular group tested the model. The `select all that apply' questions ask respondents to elaborate on their reasoning to the Likert-style questions and to suggest what the group should do next. For each of these `reasoning' and `what to do next' questions, there are between 5-10 options to choose from, and students are limited to selecting no more than three options.

The PLIC presents respondents with a hypothetical scenario where two groups of physicists are completing a mass-on-a-spring experiment to test a model where the period of oscillation of the bouncing mass is $T = 2\pi\sqrt{\frac{m}{k}}$. The first group conducts 10 repeated trials for the period of oscillation for two different masses and uses the given equation to find \textit{k} in each case. Group 2 conducts two repeated trials for the period of oscillation for 10 different masses, plots $T^{2}$ versus \textit{m} and fits to a straight line with the intercept fixed at the origin (one-parameter fit). Finally, Group 2 attempts to fit a straight line with a free intercept (two-parameter fit). In addition to the `select all that apply' questions outlined above, respondents are also asked which fit Group 2 should use and why, as well as which group they think did a better job at testing the model and why.

In scoring the PLIC, we focus on the `select all that apply' questions. That is, we do not score based on which group a student thinks did better or how well they think the group tested the model, only those ideas that they used to make their decisions. Based on responses from 24 physics post-docs, lecturers, and faculty, 1-2 responses per question that were selected by at least $50\%$ of respondents were identified as expert (E). Respondents who select \textit{at least one} of these E responses on a question receive 1 point. Further, 1-2 responses per question that were selected by more than $30\%$, but less than $50\%$ of experts, were identified as partially-expert (P). Respondents who failed to select any E responses, but selected \textit{at least one} P response, are awarded 0.5 points. Finally, 2-3 responses per question that were picked by less than $10\%$ of experts were identified as being particularly novice (N). Respondents selecting \textit{at least one} N response have 0.25 points deducted from their score. All other responses to a question have no impact on a respondent's score on that question. All scores are floored at zero, so even if a respondent selects only neutral and novice responses to a question, they still receive a score of 0 on that question.

This scoring scheme allows students to obtain a maximum possible score on a question regardless of how many responses they select; we allow students to select up to three responses, but do not penalize for picking fewer. This scheme also provides credit for partial displays of critical thinking and differentiates students who answer in correct and partially correct ways from students who still have novice ideas about physics experimentation. On any question a student can receive a score between 0 and 1 in increments of 0.25. The PLIC's current format of 10 questions then allows for a maximum possible total score of 10 points.

The original 24 experts (upon whom the scoring scheme was based) obtained an average overall score on the PLIC of $8.61\pm0.17$. We subsequently received 24 additional responses from physics experts who scored $7.9\pm0.3$. Since neither of these data are normally distributed and sample sizes are small, we used a non-parametric Mann-Whitney U-test to compare the two sample distributions. They are statistically indistinguishable ($p = 0.13$), hence we combine these data for later comparisons.

Over the course of the 2017-2018 academic year, we collected PLIC responses before and after course instruction from 25 courses across 12 institutions. A total of 2681 valid responses were collected from students who completed the survey, consented to participate in the study, and indicated they were at least 18 years of age. Of these valid responses, pre- and post-responses were matched for individual students from the student ID or full name they provided at the end of the survey. In 2017-2018, we collected matched pre- and post-surveys from 726 students. The lab level and type (traditional or SQI) for each class were inferred from information provided by instructors about their course via a course information survey (CIS). The CIS is part of an automated system associated with the PLIC, which was adapted from Ref.~\citep{wilcox2016alternative}.

\vspace{-0.4cm}
\section{Methods}

\vspace{-0.2cm}
To assess concurrent validity, we split our matched dataset by physics maturity and compared performance on the PLIC. This split dataset includes 584 students in first-year (FY) labs, 108 students in beyond-first-year (BFY) labs, and 34 students in graduate level labs. To assess the impact of SQIlab instruction on students' PLIC performance we grouped FY students according to the type of lab their instructor indicated they were running as part of the CIS. The data includes 383 students who participated in traditional FY physics labs and 201 who participated in FY SQIlabs. 
We also examined students' responses to individual questions in detail to illuminate the differences in overall performance.

{\renewcommand{\arraystretch}{0.5}
\begin{table}[htbp]
\caption{Performance on the PLIC across different levels of physics maturity. \textit{N} is the number of matched responses within a dataset, except in the case of the expert surveys where respondents only filled out the survey once. Significance levels and effect sizes are reported for differences in pre- and post-means within each group of students.\label{Score_Level}}
\begin{ruledtabular}
\begin{tabular}{cl c c c c}
& \textit{\textbf{N}} & \textbf{Pre Avg.} & \textbf{Post Avg.} & \textit{\textbf{p}} & \textit{\textbf{d}}\\ 
\hline
FY & 584 & $5.61\pm0.07$ & $5.79\pm0.07$ & 0.064 & 0.10\\
BFY & 108 & $6.34\pm0.17$ & $6.34\pm0.15$ & 0.994 & $\ll 0.01$\\
Grad & 34 & $6.9\pm0.2$ & $6.8\pm0.3$ & 0.823 & 0.05\\
Experts & 48 & $8.2\pm0.2$ & & \\
\end{tabular}
\end{ruledtabular}
\end{table}}

\begin{figure}
\includegraphics[width=0.85\linewidth]{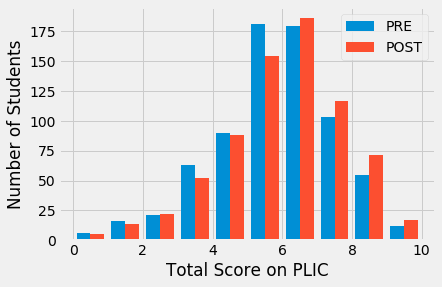}
\caption{Distributions of matched pre- and post-scores on the PLIC. $N = 726$.\label{Distribution}}
\end{figure}

Our data follows an approximately normal distribution with roughly equal variances in pre- and post-scores (see Fig.~\ref{Distribution}). For this reason, we used parametric statistical tests to compare paired (paired t-test) and unpaired sample means (unpaired t-test), and a one-way Analysis of Covariance (ANCOVA) to evaluate the effect of lab treatment on post-scores with pre-scores as a covariate. We used Cohen's \textit{d} to calculate effect sizes between two sample means and Cohen's \textit{f} to calculate effect sizes of the independent variable and covariate on the dependent variable in ANCOVA.

\vspace{-0.4cm}
\section{Results and Discussion}

\vspace{-0.2cm}
\subsection{Concurrent Validity}

\vspace{-0.2cm}
We begin by comparing respondents' performance on the PLIC by physics maturity. In Table~\ref{Score_Level}, we report the average scores for students enrolled in different level physics lab courses, as well as for our 48 experts who only took the PLIC once. The significance level and effect sizes between pre- and post-mean scores within each group are also included.

For all three groups of students, the differences between the pre- and post-instruction means are not statistically significant and the effect sizes are small or very small (Cohen's $d \leq 0.1$). Conversely, the pre-instruction means are statistically different between all groups (unpaired t-test, $p \ll 0.01$) other than between students in BFY and graduate level labs ($p = 0.12$). The effect sizes ranged from small (Cohen's $d = 0.31$) between the students in BFY and graduate labs to very large ($d = 1.6$) between students in FY labs and experts. The clear differences in means between groups of differing physics maturity, coupled with the lack of measurable increase in mean scores following instruction at any level, may imply that these differences arise from selection effects rather than cumulative instruction. This has been seen in other evaluations of students' lab sophistication as well~\citep{wilcox2017improvement}.

\begin{figure}
\includegraphics[width=0.85\linewidth]{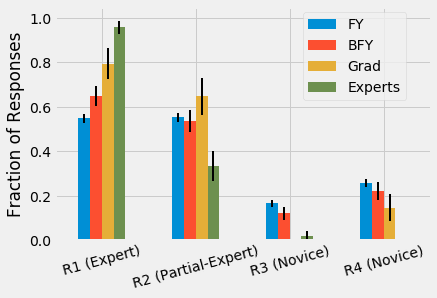}
\caption{Fraction of respondents that select a particular item in response to a question on the pre-instruction PLIC about how well the k-values from data collected using two different masses agree with each other. The responses R1-R4 are explained in the text. $N(FY) = 584$, $N(BFY) = 108$, $N(Grad) = 34$, $N(Experts) = 48$.\label{Level_Responses}}
\end{figure}

We illustrate in Fig.~\ref{Level_Responses} how the differences in physics maturity play out on one question from the pre-instruction PLIC. We use the pre-survey here since it allows us to examine students thinking before instruction, and eliminates concern about students seeing the PLIC for a second time. We chose the question presented here because it clearly demonstrates the inherent differences in item selections by respondents from these different groups. In this question, respondents were asked: `How similar or different do you think Group 1's spring constant (k) values are?' (Likert-style, not scored) and `what features were most important in comparing the two k values?' (`select all that apply', scored) The answers to the `select all that apply' question that will affect a respondent's score, as well as their classifications, are:
\begin{itemize}
\vspace{-0.25cm}
\item R1 (Expert) - the difference in the k-values compared to their uncertainties,
\vspace{-0.25cm}
\item R2 (Partial-Expert) - the size of the uncertainties (or the variability in the data),
\vspace{-0.25cm}
\item R3 (Novice) - the difference between the oscillation periods of the masses,
\vspace{-0.25cm}
\item R4 (Novice) - how they accounted for human error.
\end{itemize}

As expected, the expert answer to this question is picked increasingly more often with the physics maturity of the respondent. Additionally, the two novice answers to this question are picked less frequently with additional physics maturity of the respondent. The partial-expert response, however, is picked almost equally as often by all students. The fact that students equally value a partially correct answer 
{\renewcommand{\arraystretch}{0.5}
\begin{table}[htbp]
\caption{Table of ANCOVA results for post-instruction scores.\label{ANCOVA}}
\begin{ruledtabular}
\begin{tabular}{cl c c c c}
& \textbf{Effect} & \textit{\textbf{F}} & \textbf{Effect Size} & \textit{\textbf{p}} \\ 
\hline
Post-scores & Pre-scores & 18.39 & 0.18 & $\ll 0.01$ \\
& Lab Treatment & 18.61 & 0.18 & $\ll 0.01$ \\
\end{tabular}
\end{ruledtabular}
\end{table}}
at all levels could indicate that most intro to advanced physics lab courses have some focus on uncertainty. Though we have described just one question in detail here, the performance differences are present across all of the questions on the PLIC. Experts score higher than students in BFY labs who, in turn, score higher than students in FY labs. As seen here, these performance differences are manifest in the expertness of responses.

\vspace{-0.4cm}
\subsection{Performance by Lab Type}

\begin{figure}
\includegraphics[width=0.85\linewidth]{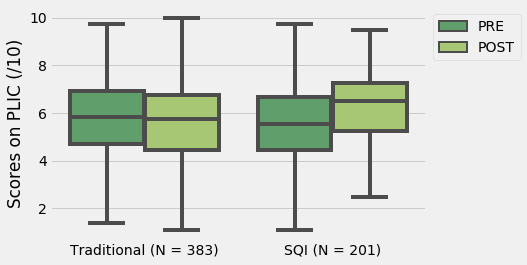}
\caption{Box plots of overall PLIC scores grouped by the type of lab students participated in.\label{Lab_Overall}}
\end{figure}

\vspace{-0.2cm}
We now examine how students participating in labs designed to meet the instructional goals of the PLIC performed in comparison to their counterparts in traditional lab settings. We limit our analysis to FY labs where we had three labs that were best described as being of the SQI format, while the rest were of the traditional format. The overall performance of these two groups of students is illustrated in Fig.~\ref{Lab_Overall}.

We performed an ANCOVA comparing PLIC post-scores across lab treatments using pre-scores as a covariate (Table~\ref{ANCOVA}). We see that, controlling for pre-instruction scores, lab treatment does have a statistically significant impact on post-instruction scores with a medium effect.

\begin{figure}
\includegraphics[width=0.85\linewidth]{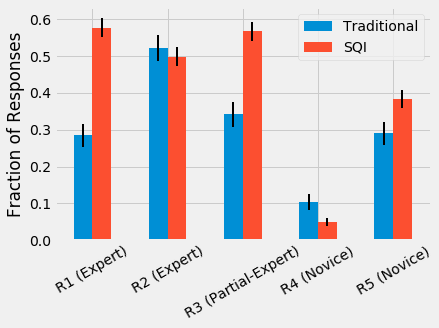}
\caption{Fraction of students that select a particular item in response to a question on the PLIC post-survey about how well data fits a 2-parameter best fit line. The responses R1-R5 are explained in the text. $N(Traditional) = 383$, $N(SQI) = 201$.
\label{Lab_Distribution}}
\end{figure}

To see how these differences in overall scores arise, we looked at how students in the two lab types compare in their responses to one question. Fig.~\ref{Lab_Distribution} shows the fraction of students who selected particular responses to one of the questions on the PLIC post-survey grouped by lab type. We show only the post-instruction data since there was no statistically significant difference in pre-instruction scores on this question between the two groups. Again, we chose this question because it most clearly illustrates key differences in student thinking by respondents from the two lab types.

In this question, students are asked: `How similar or different do you think Group 2's data are from the new best-fit line?' (Likert-style, not scored) and `what features were most important in comparing the fit to the data?' (`select all that apply', scored) The responses that will affect a respondent's score are:
\begin{itemize}
\vspace{-0.25cm}
\item R1 (Expert) - the way points are scattered above and below the line,
\vspace{-0.25cm}
\item R2 (Expert) - how close the points are to the line compared to their uncertainties,
\vspace{-0.25cm}
\item R3 (Partial-Expert) - number of points with uncertainties crossing the line,
\vspace{-0.25cm}
\item R4 (Novice) - the number of outliers,
\vspace{-0.25cm}
\item R5 (Novice) - number of points above the line compared to the number below.
\end{itemize}

One of the expert responses, R2, is a common response by students in both lab types and is mostly unaffected by instruction. However, students in SQIlabs favored the other expert response, R1, much more than students in traditional labs following instruction. Both R1 and R5 are concerned with the overall distribution of points about the best fit line. Thus, students taught in SQIlabs appear to become more engaged with the importance of the distribution of residuals, albeit some students may oversimplify this to the sheer number of points above and below the line. Nonetheless, traditionally taught students maintain their interest in the closeness of the points to the best fit line and do not acknowledge the importance of the distribution of residuals any more than they did prior to instruction. Interestingly, students in both labs become less interested in the presence of outliers following instruction. Again, though we have included only one question here for illustration purposes, students in SQIlabs see, on average, increases in performance on all but one question, whereas the average scores for students in traditional labs increase on only 4 out of 10 questions.

\vspace{-0.4cm}
\section{Conclusions}

\vspace{-0.2cm}
With the need for large-scale instructional reform in physics lab instruction, there will be an equally important need for a method to evaluate these new instructional goals. Here, we have introduced one such method for measuring these goals, the PLIC. By comparing the performance of physics experts to students in FY and BFY undergraduate labs, and those in graduate-level labs, we have demonstrated the concurrent validity of the assessment. Though likely due to selection effects rather than instruction, respondents with a greater level of physics maturity perform consistently better on the PLIC.

Further, though we have seen no statistically significant shifts in performance following instruction for large cohorts of students, this is not the case for students enrolled in SQIlabs designed to teach the skills that the PLIC is designed to measure. Using ANCOVA, we have shown that SQIlabs have a statistically significant and medium effect on PLIC performance. Despite there being no measurable difference in PLIC performance before instruction, students in SQIlabs had created separation from their counterparts in traditional labs, in terms of PLIC performance, following instruction. These studies establish one measure of validity of the PLIC and its importance in measuring the skills that we aim to teach in meeting the new guidelines for physics labs in the future.

\vspace{-0.4cm}
\acknowledgments

\vspace{-0.2cm}
This material is based upon work supported by the National Science Foundation under Grant No. 1611482. We would like to acknowledge Carl Wieman, members of CPERL for their useful feedback, and Heather Lewandowski and Bethany Wilcox for their support developing the administration system.


\begin{thebibliography}{9}
\bibitem{millar2004role}R.~Millar, \emph{The role of practical work in the teaching and learning of science} (National Academy of Sciences, Washington, D.C., 2004).
\bibitem{hofstein2004laboratory}A.~Hoffstein \& V.N.~Lunetta, \emph{Sci. Educ.} \textbf{88}, 1 (2004).
\bibitem{holmes2017value}N.~Holmes, \emph{et al.}, \emph{Phys. Rev. Phys. Educ. Res.} \textbf{13}, 010129 (2017).
\bibitem{holmes2015teaching}N.G.~Holmes, \emph{et al.}, \emph{PNAS.} \textbf{112}, 36 (2015).
\bibitem{etkina2006using}E.~Etkina, \emph{et al.}, \emph{Am. J. Phys.} \textbf{74}, 979 (2006).
\bibitem{kozminski2014aapt}Kozminski, Joseph, \emph{et al}. \emph{AAPT Recommendations for the Undergraduate Physics Laboratory Curriculum.} (2014).
\bibitem{wilcox2016open}B.R.~Wilcox and H.~Lewandowski. \emph{Phys. Rev. Phys. Educ. Res.} \textbf{12}, 020132 (2016).
\bibitem{adams2011development}W.K.~Adams \& C.E.~Wieman, \emph{Int. J. Sci. Educ.} \textbf{33}, 9 (2011).
\bibitem{madsen2017resource}A.~Madsen \emph{et al.} \emph{Am. J. Phys.} (2017).
\bibitem{HolmesPERC2015} N.G.~Holmes \& C.E.~Wieman, in \emph{PERC proceedings, Sacramento, CA, 2016}.
\bibitem{quinn2018interview}K.N.~Quinn \emph{et al.}, in \emph{PERC proceedings, Cincinnati, OH, 2017}.
\bibitem{wilcox2016alternative}B.R.~Wilcox, \emph{et al.}, \emph{Phys. Rev. Phys. Educ. Res.} \textbf{12}, 010139 (2016).
\bibitem{wilcox2017improvement}B.R.~Wilcox \& H. Lewandowski, \emph{Phys. Rev. Phys. Educ. Res.} \textbf{13}, 023101 (2017).
\end{thebibliography}
\end{document}